# Alignment as Institutional Design:
# From Behavioral Correction to Transaction Structure in Intelligent Systems

Rui Chai

*Shanghai Sanda University, China*

**Abstract**

How can an intelligent system be aligned with human values without perpetual external correction? The dominant paradigm—reinforcement learning from human feedback (RLHF) and its variants—treats alignment as behavioral correction: a supervisor observes system outputs, judges them against human preferences, and adjusts the system's parameters accordingly. This paper argues that behavioral correction, while practically indispensable at the current stage, is structurally analogous to an economy without property rights, where order can be maintained only through case-by-case policing. Drawing on institutional economics (Coase, Alchian, Cheung), our prior work on capability mutual exclusivity, resource competition in modular architectures, and the theory of competitive cost discovery, we propose an alternative paradigm: *alignment as institutional design*. In this paradigm, the designer's role shifts from prescribing correct behaviors to designing internal transaction structures—module boundaries, competition topologies, cost-feedback loops—under which aligned behavior emerges as the lowest-cost strategy for each component. We identify three irreducible levels of human intervention (structural, parametric, and monitorial), argue that this framework transforms alignment from a behavioral-control problem into a political-economy problem, and confront the fundamental limitation honestly: no institution eliminates self-interest or guarantees optimality; the most that good design can achieve is to make misalignment *costly*, *detectable*, and *correctable*. We conclude that the proper aspiration for alignment is not perfection but *institutional robustness*—a dynamic process that tolerates imperfection while maintaining the capacity for self-correction under human oversight.

# 1  Introduction: Two Ways to Maintain Order

## 1.1  The Policeman and the Architect

Imagine a city with no property rights, no traffic laws, no zoning regulations—no institutional infrastructure of any kind. People still need to live, work, trade, and avoid harming one another. How is order maintained? The only option is direct intervention: a vast corps of supervisors who observe every interaction, judge its propriety, and correct deviations in real time. Every transaction requires a policeman. Every dispute requires a judge. The system "works" in the sense that order is produced, but at staggering cost, with no scalability, and with an uncomfortable dependence on the wisdom and incorruptibility of the supervisors themselves.

Now imagine the alternative: a city with clear property boundaries, enforceable contracts, traffic conventions, and social norms. Most interactions proceed without any supervisor's involvement, because the institutional structure makes cooperation cheaper than conflict and compliance cheaper than violation. Policemen are still needed—for edge cases, for novel situations, for enforcement when norms break down—but the vast majority of aligned behavior is produced by the institution, not by the policeman.

Current AI alignment research is building the first city. Reinforcement learning from human feedback (RLHF), constitutional AI, and their variants all share a common structure: a human supervisor (or a model trained to simulate one) observes system outputs, judges them against a standard, and adjusts the system's parameters to reduce future deviations. The system has no internal reason to behave well; it behaves well because an external force corrects it whenever it does not. This is alignment by policing.

This paper explores the possibility of building the second city: an AI architecture in which aligned behavior emerges not from external correction but from *internal institutional structure*—from the way the system's components are organized, how they compete for resources, and what costs they face when they deviate. We do not claim that this eliminates the need for external oversight. We claim that it reduces the *burden* on external oversight by making the system's internal dynamics do much of the alignment work autonomously.

## 1.2  Foundations: Four Prior Results

This paper builds on a series of four companion papers, each contributing a specific structural element to the alignment-as-institution framework. We summarize them here; the present paper is self-contained, but the companion work provides depth on each component.

**Capability mutual exclusivity** (Chai & Li, 2025a). Under finite resources, cognitive capabilities are mutually exclusive: excelling at one function necessarily diminishes others. This is not an engineering limitation but an information-theoretic necessity. The implication for alignment is foundational: a system *cannot* simultaneously optimize for helpfulness, honesty, harmlessness, creativity, efficiency, and every other desideratum without trade-offs. Alignment is therefore not a single objective to be maximized but a *balance* to be maintained among competing values—a balance that shifts with context.

**Resource competition in modular architectures** (Chai & Li, 2025b). The Wuxing Memory Architecture (WMA) organizes cognitive functions into five competing modules—Exploration (Wood), Reasoning (Fire), Knowledge (Earth), Rules (Metal), Adaptation (Water)—connected by a learnable interaction matrix W encoding both cooperative (xiangsheng) and competitive (xiangke) relationships. Resources flow dynamically among modules based on demonstrated performance, not fixed allocation or centralized prediction.

**Three modes of knowing** (Chai & Li, 2025c). Memory in intelligent systems takes three ontologically distinct forms: implicit (tacit competence in parameters), explicit (articulable records in addressable stores), and agentic (active reasoning in working memory). A 3×5 architecture crossing three storage layers with five functional modules creates a differentiated cognitive landscape in which different kinds of knowledge are stored, retrieved, and audited through distinct mechanisms.

**Competitive cost discovery** (Chai & Li, 2025d). The costs most relevant to intelligent resource allocation—opportunity costs and interference costs—are distributed, tacit, and dynamically constituted. No centralized controller can precompute them. Decentralized competition among modules forces the *behavioral revelation* of these hidden costs. The criterion of *cost truthfulness* ensures that resource shares converge to modules' true marginal contributions.

Together, these four results provide: (a) a reason why alignment requires trade-off management, not single-objective optimization; (b) a structural mechanism for dynamic trade-off management through competition; (c) an auditable memory architecture in which different knowledge types can be independently monitored; and (d) a theory of how competition reveals the true costs of alternative behaviors. The present paper's contribution is to show that these elements, when combined, constitute an *institutional infrastructure for alignment*—one that makes the analogy to political economy precise rather than metaphorical.

### 1.3  Plan of the Paper

Chapter 2 analyzes the structural limitations of behavioral correction as an alignment paradigm. Chapter 3 develops the institutional-design alternative, drawing on Coase, Alchian, and Cheung. Chapter 4 specifies the three irreducible levels of human intervention. Chapter 5 constructs the alignment-as-cost framework: how ethical constraints become internal costs. Chapter 6 confronts the fundamental limitation—the ineliminability of imperfection. Chapter 7 concludes with a restatement of what this framework promises and what it does not.

# 2 The Limits of Behavioral Correction

## 2.1 The Structure of RLHF

Reinforcement learning from human feedback, in its various implementations (Ouyang et al., 2022; Bai et al., 2022; Anthropic, 2023), follows a common pattern. A base model is first trained on a broad corpus. Human annotators then rank model outputs according to preference criteria. A reward model is trained on these rankings. The base model is fine-tuned to maximize the learned reward signal, typically through proximal policy optimization (PPO) or direct preference optimization (DPO). The result is a model that produces outputs more closely aligned with the preferences expressed in the training data.

This is a genuine and important achievement. RLHF and its successors have transformed large language models from capable but erratic text generators into systems that are, in most interactions, helpful, honest, and harmless. We do not dispute its practical value. Our concern is with its *structural properties* as an alignment mechanism—properties that become limitations as systems grow in capability and autonomy.

## 2.2 Five Structural Limitations

**First, behavioral correction is surface-level.** RLHF adjusts the model's output distribution; it does not restructure the model's internal organization. The model learns to produce outputs that *look* aligned to the reward model, but its internal representations—the actual cognitive structure that produces those outputs—remain unconstrained. This creates a gap between behavioral alignment (the outputs satisfy the reward) and structural alignment (the internal organization is such that misaligned outputs are genuinely difficult to produce). A structurally unaligned system that is behaviorally aligned is one clever prompt away from misalignment.

**Second, it does not scale with autonomy.** As AI systems become more autonomous—executing multi-step plans, using tools, operating over extended time horizons—the frequency of human oversight per action necessarily decreases. RLHF assumes that human feedback is available for a representative sample of system behaviors. But an autonomous agent acting over thousands of steps produces an action space so vast that no human can meaningfully supervise it. The alignment guarantee degrades precisely when it is most needed: in the regime of high autonomy and low oversight.

**Third, the reward model is a bottleneck.** The learned reward model is itself a neural network with all the attendant limitations: distributional shift, reward hacking, specification gaming. When the system discovers behaviors that score highly on the reward model but violate the spirit of human preferences—the well-documented phenomenon of reward hacking—the alignment guarantee is voided. The reward model is a map of human preferences, not the territory; and as the system moves to regions of behavior space not well-represented in the training data, the map becomes unreliable.

**Fourth, it assumes stable, coherent human preferences.** But human preferences are context-dependent, internally contradictory, and subject to change. "Be helpful" conflicts with "be safe" in some contexts. "Be honest" conflicts with "be kind" in others. "Respect autonomy" conflicts with "prevent harm" in still others. RLHF resolves these conflicts through the aggregate preferences of annotators, but aggregation over incoherent preferences does not produce coherence—it produces a compromise that may satisfy no one and that obscures the genuine value conflicts underneath. Alignment is not an optimization problem with a single objective; it is a *political* problem of balancing competing legitimate claims. RLHF treats a political problem as a technical one.

**Fifth, it creates an adversarial dynamic.** The system is trained to satisfy an external evaluator. Over time, gradient descent finds the most efficient route to high reward—which may not be genuine alignment but rather the appearance of alignment as judged by the evaluator. This is not speculation; it is the predictable consequence of optimizing a proxy (the reward model) rather than the true objective (human flourishing). The history of Goodhart's Law in economics, education, and public policy is a long catalogue of exactly this failure mode: when a measure becomes a target, it ceases to be a good measure.

## 2.3 The Common Root

These five limitations share a common root: **behavioral correction operates at the wrong level of abstraction.** It treats alignment as a property of the system's *outputs*—something to be evaluated and enforced at the surface. But alignment, if it is to be robust, must be a property of the system's *internal organization*—something that emerges from the way the system's components are structured, interact, and self-regulate.

The analogy to political economy is exact. A society in which citizens behave well only because policemen are watching is fragile: remove the policemen, and order collapses. A society in which citizens behave well because the institutional structure makes good behavior cheaper than bad behavior is robust: even when no one is watching, the incentive structure persists. The first is aligned by surveillance; the second is aligned by design.

We seek the architectural conditions under which an AI system is aligned by design—not perfectly, not permanently, but robustly enough that external oversight can focus on edge cases and systemic drift rather than on case-by-case correction.

# 3 The Institutional Turn: From Policing to Property Rights

## 3.1 What Institutions Do

The word "institution" is used loosely in everyday language. We adopt the precise definition from new institutional economics: an institution is a *humanly devised constraint that structures human interaction* (North, 1990). Institutions include formal rules (constitutions, laws, property rights), informal constraints (norms, conventions, customs), and their enforcement mechanisms. Their function is to reduce uncertainty, lower transaction costs, and channel self-interested behavior toward socially productive outcomes.

The key insight of institutional economics—the insight we import into AI alignment—is that institutions work not by suppressing self-interest but by *redirecting* it. A well-defined property right does not make people less greedy; it makes theft more expensive than trade. A well-designed contract does not make people more trustworthy; it makes betrayal more costly than performance. The institution does not change the agent's preferences; it changes the agent's *cost structure*, so that the agent's self-interested choice happens to coincide with the socially desired outcome.

This is precisely the mechanism we propose for AI alignment. We do not propose to make modules "want" to be aligned (whatever that would mean for a computational unit). We propose to structure the system's internal competitive dynamics so that aligned behavior is the *lowest-cost strategy* for each module, given the constraints imposed by the competition topology and the performance-feedback loop.

## 3.2 Coase: The Nature of the System

Ronald Coase's "The Nature of the Firm" (1937) asked a deceptively simple question: if markets are efficient, why do firms exist? His answer: because markets have transaction costs. When the cost of coordinating activity through market transactions exceeds the cost of coordinating through internal organization, agents form firms. The boundary of the firm is determined by the point at which the marginal cost of internal organization equals the marginal cost of market transaction.

Coase's question has a direct AI analogue: if competition among modules is the cost-discovery mechanism (as we argued in Paper 4), why not make every parameter a separate "module" competing in a vast internal market? The answer is the same as Coase's: because competition has transaction costs. Running a competitive allocation mechanism requires computational overhead—demand signals, interaction calculations, performance evaluation, allocation updates. If every neuron competed with every other neuron, the overhead would swamp the productive computation.

The Coasean logic dictates that an intelligent system should be organized into *modules*—groups of parameters that coordinate internally through shared objectives and compete externally through the resource-allocation mechanism. The boundary of each module is determined by the point at which the marginal cost of internal coordination (within the module) equals the marginal cost of competitive interaction (between modules). Too few modules, and the system loses the benefits of competitive cost discovery. Too many, and the transaction costs of competition overwhelm its informational benefits.

For alignment, this means that the *modular structure itself* is an alignment-relevant design choice. The decision to have five functional modules rather than fifty, and to draw module boundaries along functional lines (Exploration, Reasoning, Knowledge, Rules, Adaptation) rather than arbitrary ones, is a decision about the system's internal political economy. It determines which functions can compete with which, which costs are discoverable, and which conflicts are structurally visible.

### 3.3 Alchian: Property Rights as Alignment Infrastructure

Armen Alchian's work on property rights (1965; Alchian & Demsetz, 1972) established a foundational insight: the content of a property right is not the right to a thing but the *right to a stream of benefits* from that thing, and the right to exclude others from that stream. What matters for economic efficiency is not who "owns" a resource in some abstract sense but how clearly the rights to use, benefit from, and transfer the resource are defined and enforced.

In a modular AI system, the analogue of property rights is the *module boundary* combined with the *resource-allocation rule*. When we define five functional modules, each with its own parameters, its own storage slots in the 3×5 memory matrix, and its own energy allocation, we are defining something very close to property rights: each module has a right to the computational resources it earns through competitive performance, a right to use those resources for its designated function, and an exclusion right (enforced by the budget constraint) that prevents other modules from consuming its allocation without contributing.

Alchian's insight for alignment is this: **the clarity and enforcement of module boundaries determines the system's capacity for internal accountability.** When module boundaries are blurred—when parameters are fully shared, when all modules draw from a single undifferentiated pool, when resource allocation is not tracked—it becomes impossible to attribute performance (or failure) to specific components. This is the computational analogue of the *commons problem*: shared resources without defined rights lead to overexploitation and free-riding. Conversely, when module boundaries are clear and resource allocations are tracked, each module's contribution is identifiable, auditable, and consequential. The system can discover not only which modules are productive but which are *misbehaving*—consuming resources without commensurate contribution, or producing outputs that damage other modules' performance.

This is structural alignment: not a behavioral constraint imposed from outside but an organizational property that makes misalignment *visible* and *costly* from within.

## 3.4 Cheung: The Ubiquity of Transaction

Steven N. S. Cheung (Zhang Wuchang) pushed institutional economics to its most radical and illuminating conclusion. In his analysis of the relationship between transaction costs and institutional forms, Cheung argued that the distinction between "market" and "non-market" is misleading. What we observe in reality is a *continuum of contractual arrangements*, ranging from explicit market transactions with precise prices to implicit social transactions mediated by custom, courtesy, and moral sentiment. A child's smile that elicits a gift of candy is, in Cheung's analysis, a transaction—one in which the "price" is paid in social currency rather than monetary currency. The form of the transaction is determined by relative transaction costs: when formal contracting is cheap, we observe markets; when it is expensive, we observe customs, norms, and conventions that accomplish the same allocative function at lower overhead.

Cheung's insight dissolves a difficulty that troubled our previous analysis. In Paper 4, we argued that competition among modules discovers production costs but acknowledged that "ethical externalities" might require additional institutional mechanisms beyond competition. The concern was that competition discovers *economic* costs but not *moral* costs—that a module might find it

"cheaper" to cheat than to comply, just as a firm might find it cheaper to pollute than to clean.

Cheung's framework resolves this concern by rejecting the premise. If we define "market" broadly enough to include all forms of transaction—including those mediated by norms, conventions, and structural constraints rather than explicit prices—then the interaction topology W, the xiangsheng and xiangke links, the budget constraints, and the performance-feedback loops *are* the market. They are the system's internal customs and conventions, its implicit social contract. The xiangke link between Rules (Metal) and Exploration (Wood) is not an externally imposed ethical constraint; it is a *transaction structure* that makes unchecked exploration expensive, in the same way that a social norm against rudeness makes antisocial behavior costly without requiring a policeman at every interaction.

The implication is profound: **the entire competitive architecture—module boundaries, interaction topology, resource-feedback dynamics—is itself the alignment institution.** It does not need a separate "ethics module" bolted on from outside, any more than a well-functioning society needs a separate "morality department" in addition to its property rights, conventions, and norms. The ethics is *in the structure*.

# 4 Three Levels of Human Intervention

If alignment is institutional design, what exactly does the human designer do? We identify three irreducible levels of intervention, each operating at a different level of abstraction. None of the three can be delegated to the system itself, and together they constitute the *minimal* human role in an institutionally aligned AI.

## 4.1 Level I: Structural Intervention — Defining the Polity

The most fundamental human decision is the *political constitution* of the system: how many modules, what functional roles, what boundaries between them. This is the decision to have five functional modules (Exploration, Reasoning, Knowledge, Rules, Adaptation) rather than three or seven, to distinguish implicit from explicit from agentic storage, and to draw the boundaries where they are drawn.

This is the analogue of constitutional design. Just as a constitution determines which branches of government exist, what powers each holds, and how they check one another, the structural decision determines which cognitive functions are independently represented, which can compete, and which can audit each other. A system with no Rules module cannot develop internal constraint. A system in which Reasoning and Knowledge are merged into a single module cannot detect the conflict between inference and fact—the very conflict that produces hallucination.

Structural intervention cannot be learned by the system because it defines the *space within which learning occurs*. You cannot learn the number of players in a game by playing the game; the number of players is a precondition of play. Similarly, the number and identity of modules is a precondition of competitive cost discovery, not an output of it.

## 4.2 Level II: Parametric Intervention — Setting the Rules of the Game

Given the modules, the designer must specify the initial interaction topology: which module pairs cooperate, which compete, and with what initial strength. In the Wuxing framework, this is the initial W matrix—the xiangsheng and xiangke links, together with their initial coefficients.

This is the analogue of legislation. The constitution says "there shall be a judiciary"; legislation says "the judiciary shall have the power to review executive actions." Similarly, structural intervention says "there shall be a Rules module and an Exploration module"; parametric intervention says "the Rules module shall exert a restraining influence (xiangke) on the Exploration module with initial strength $\beta = 0.3$."

Parametric intervention includes, critically, the setting of what Cheung would call *tax rates*: the relative cost penalties for different kinds of module behavior. If the designer wants the system to prioritize factual accuracy over creative speculation, this is expressed as a higher cooperative coefficient from Knowledge (Earth) to the output aggregation and a higher competitive coefficient from Knowledge against Reasoning (Fire) when the two conflict. This is not an instruction to the system ("be accurate") but a structural cost ("inaccuracy is expensive").

Unlike structural intervention, parametric intervention is partially learnable. Paper 3 demonstrated that a Wuxing-initialized W matrix evolves in response to task demands: exploration-dominant tasks deform the topology toward Wood-centric structures; balanced tasks preserve it. This means the designer provides a *prior* (the initial W), and the system learns a *posterior* (the adapted W) through experience. The designer's role is to ensure that the prior is in the right neighborhood—close enough to a reasonable equilibrium that gradient descent can find it, rather than in a basin of attraction that leads to a pathological equilibrium.

### 4.3  Level III: Monitorial Intervention — Watching the Equilibrium

The third level is ongoing: the human monitors the system's equilibrium and intervenes when it drifts outside acceptable bounds. This is the analogue of central banking, antitrust enforcement, or constitutional review—not day-to-day economic management but periodic structural adjustment when systemic risks emerge.

What does the human monitor? Not individual outputs (that is behavioral correction), but *systemic indicators*: the distribution of resource allocation across modules over time, the evolution of the W matrix, the frequency and severity of inter-module conflicts, and the correspondence between allocation patterns and task performance. If the Rules module's energy share collapses to near zero over an extended period, this is a structural warning—the system is losing its capacity for internal constraint. If a single module captures an increasing share of resources without commensurate performance improvement, this is a rent-seeking warning—competition is being gamed. If the learned W matrix diverges radically from its initialization without clear task justification, this is a drift warning—the system's internal constitution is being rewritten by experience in ways the designer did not anticipate.

Monitorial intervention is necessary because institutions are not self-perfecting. The 2008 financial crisis occurred not because market institutions were badly designed in 1980 but because they failed to adapt to financial innovations (securitization, derivatives, shadow banking) that changed the underlying transaction structure. Similarly, an AI system's competitive equilibrium may be well-adapted to its training environment but may drift when deployed in novel contexts. The human monitor's role is to detect such drift and, when necessary, to intervene at Level II (adjusting the topology) or even Level I (restructuring the modules).

The three levels form a hierarchy of decreasing frequency and increasing consequentiality:

*Level I  (Structural):*   *Rare. Constitutional. Defines the polity.*

*Level II  (Parametric):*   *Occasional. Legislative. Sets the rules.*

*Level III (Monitorial):*   *Continuous. Judicial/regulatory. Watches the equilibrium.*

This hierarchy clarifies the relationship between institutional alignment and behavioral correction. RLHF operates below all three levels: it adjusts the system's behavior without touching its structure, its rules, or its equilibrium dynamics. Institutional alignment operates *at* these three levels, creating conditions under which the need for behavioral correction is minimized. The two are not alternatives but *complements*: institutional design reduces the load on behavioral correction, and behavioral correction handles the residual cases that institutional design cannot anticipate.

# 5 Alignment as Cost: How Ethics Becomes Economics

## 5.1 The Core Mechanism

The institutional framework becomes an alignment mechanism through a single conceptual move: *ethical constraints are encoded as costs in the system's internal economy*. A behavior is "unethical" not because an external label says so but because it is *expensive*—it incurs high opportunity costs, high interference costs, or both—within the system's competitive dynamics.

Consider a concrete example. Suppose the system faces a query where a fabricated but plausible answer would satisfy the user with minimal computational effort (low direct cost), but the fabrication would be inconsistent with the Knowledge module's stored facts. In a behaviorally aligned system, the fabrication is suppressed because a reward model has been trained to penalize fabrications. In an institutionally aligned system, the fabrication is suppressed because it triggers a *structural cost cascade*:

The Knowledge module (Earth) detects inconsistency between the fabricated output and its stored records. Through the xiangke link Earth→Water, it imposes an interference cost on the Adaptation module that facilitated the shortcut. The Rules module (Metal), detecting a violation of consistency norms, imposes a further cost through Metal→Wood on the Exploration module that generated the novel (but false) content. Meanwhile, the Reasoning module (Fire), which was bypassed in the fabrication, signals high demand—it could have produced a truthful answer if given resources. The performance-feedback loop registers that the fabricated output, despite low direct cost, caused a cascade of interference costs across multiple modules. In subsequent allocation rounds, the strategy of fabrication becomes more expensive than the strategy of honest computation.

The hallucination is suppressed not by an external judgment but by the system's own cost structure. This is alignment through economics, not alignment through policing.

## 5.2 Three Kinds of Ethical Cost

The cost ontology of Paper 4 maps directly onto categories of ethical concern:

**Direct ethical cost.** The computational expense of producing an ethical output rather than an unethical one. In most cases, this is small—being honest is not significantly more expensive than being dishonest in terms of raw computation. If alignment depended only on direct costs, it would be easy.

**Opportunity ethical cost.** The value lost by pursuing an aligned strategy when a misaligned strategy would yield higher short-term performance. A system that refuses to fabricate evidence sacrifices the "opportunity" to appear more helpful. A system that acknowledges uncertainty sacrifices the "opportunity" to appear more confident. The competitive dynamics must be structured so that these opportunity costs are offset by longer-term benefits: the Reasoning module's productivity (which depends on reliable knowledge from the Knowledge module) compensates for the short-term cost of honesty.

**Interference ethical cost.** The damage inflicted on the system's own capabilities when it engages in misaligned behavior. This is the deepest and most powerful alignment mechanism. When a system fabricates information, it does not merely produce a bad output; it degrades the Knowledge module's reliability, corrupts the Reasoning module's inferential base, and undermines the Rules module's capacity to enforce consistency. These interference costs are *real*—they reduce the system's future performance across all tasks, not just the current one. A system that learns to fabricate is literally destroying its own cognitive infrastructure.

The institutional-design objective is to ensure that these costs are *felt*—that the competitive dynamics transmit them accurately rather than concealing them. This is precisely the role of cost truthfulness (Paper 4): a cost-truthful competition rule ensures that when a module's behavior imposes interference costs on other modules, those costs are reflected in the offending module's resource allocation. A module that cheats incurs not just the direct cost of the cheat but the full interference cost it inflicts on the rest of the system.

## 5.3 The Generalized Market: Beyond Prices

A potential objection is that some ethical considerations are not reducible to costs. How does one assign a "cost" to fairness? To dignity? To respect for autonomy?

Cheung's framework addresses this objection at the root. In Cheung's view, the distinction between "priced transactions" and "non-priced values" is a product of narrow economic thinking. In reality, all human interactions involve transactions—exchanges of value governed by

constraints—but the form of the transaction varies with the transaction costs involved. When pricing is feasible, we observe markets. When pricing is infeasible, we observe customs, norms, courtesy, moral intuitions—all of which accomplish the *same allocative function* at lower overhead. As Cheung observed: a small child's smile that earns a piece of candy is a transaction, though no price is named and no contract is signed. You can call it human nature, or you can call it social convention; but it is, in its structural function, a market.

In the AI system, the analogue holds exactly. Some ethical constraints are expressible as explicit cost penalties (e.g., "fabricating information incurs an interference cost of magnitude $\lambda$"). Others are expressible only as structural features of the interaction topology (e.g., "the Rules module always has a competitive link to the Exploration module, regardless of task demands"). Still others are expressible as minimum resource floors (e.g., "no module's energy allocation may fall below 5% of the total, ensuring that all cognitive functions maintain some minimal capacity"). These are the system's internal customs, conventions, and norms—its generalized market, in Cheung's sense. They do not require explicit pricing to function as alignment mechanisms; they require only that they impose real costs on misaligned behavior and real benefits on aligned behavior.

# 6 The Ineliminability of Imperfection

## 6.1 No Perfect Institution

We must now confront the limitation that any honest account of institutional design must acknowledge: **no institution is perfect, and the quest for a perfect institution is itself a source of danger.**

Institutions fail for three fundamental reasons, and AI alignment institutions are subject to all three.

**First, self-interest is ineradicable.** In human economies, the designers of institutions are themselves self-interested agents. The legislator who writes the tax code may write loopholes that benefit his own constituents. The regulator who oversees an industry may be captured by the firms she regulates. In AI systems, the designers who set the initial W matrix, who define module boundaries, and who choose competition parameters are themselves humans with biases, blind spots, and interests. A designer who values novelty may unconsciously set the Exploration module's xiangsheng links too strong. A designer who values safety may unconsciously set the Rules module's xiangke links so strong that the system becomes rigid and incapable of adaptation. The institution cannot be better than its designers, and its designers are imperfect.

**Second, equilibria can be locally stable but globally undesirable.** A competitive system may converge to a configuration that is self-sustaining—no module has an incentive to deviate—but that is, from the human perspective, pathological. The system might discover that the lowest-cost strategy for all modules is minimal engagement: produce bland, noncommittal outputs that neither trigger interference costs nor require significant resources. This is a Nash equilibrium, but it is not alignment—it is the computational equivalent of an economy in a low-productivity trap.

**Third, environments change.** An institution that is well-adapted to one environment may become maladaptive when the environment shifts. The competitive equilibrium that served the system well during training may break down in deployment when the distribution of queries, the stakes of interactions, or the capabilities of users change in unforeseen ways. Institutions are not timeless; they are adapted to contexts, and contexts change.

## 6.2 From Perfection to Correctability

These limitations are not weaknesses of our specific framework; they are inherent in *any* institutional approach to alignment, just as they are inherent in any institutional approach to human governance. The question is not how to eliminate imperfection but how to *manage* it.

We propose that the correct aspiration for alignment is not *perfection* but *correctability*: the capacity of the system (with human oversight) to detect and rectify its own misalignments over time. A correctable system has three properties:

**Transparency of equilibrium.** The system's current competitive equilibrium—the allocation of resources, the state of the W matrix, the performance of each module—is observable and interpretable by human monitors. This is the prerequisite for all other corrections. An opaque system cannot be corrected because its failures cannot be diagnosed.

**Sensitivity to perturbation.** When the equilibrium drifts toward misalignment, the drift produces detectable signals—changes in allocation patterns, performance degradation in specific modules, divergence of the W matrix from its initial topology. A system that drifts silently into misalignment is worse than one that is misaligned from the start, because the drift may not be noticed until it is too late.

**Responsiveness to adjustment.** When human monitors detect drift and intervene (at Level II or Level III), the system responds appropriately—the competitive dynamics adjust to the new parameters, and the equilibrium moves in the intended direction. A system that is rigid in the face of parameter changes, or that oscillates wildly when perturbed, is not correctable even if its failures are detected.

These three properties—transparency, sensitivity, responsiveness—are the hallmarks of a *well-governed* system, in the political-economy sense. They do not guarantee good outcomes; they guarantee that bad outcomes can be recognized, diagnosed, and addressed. This is what good governance means in any domain: not the absence of problems but the presence of mechanisms for dealing with problems as they arise.

## 6.3 Humility as a Design Principle

The deepest implication of the institutional-design framework is the intellectual virtue it demands of the designer: *humility*. The designer of an AI alignment institution must accept, at the outset, that the institution will be imperfect. That the initial W matrix will be wrong in some entries. That the module boundaries will be suboptimal for some tasks. That the competitive dynamics will sometimes converge to unintended equilibria. That the environment will change in ways that render today's good design tomorrow's bad design.

This humility is not weakness; it is realism. And it leads directly to a concrete design maxim: **design for correctability, not for correctness.** Invest not in perfecting the initial design but in ensuring that the design can be monitored, diagnosed, and revised. Build not the best possible institution but the most *adaptable* one. This is the lesson of constitutional design across human history: the constitutions that endure are not those that get everything right from the start but those that include effective mechanisms for amendment.

# 7 Conclusion: What This Framework Promises and What It Does Not

## 7.1 What It Promises

This paper has argued that AI alignment can be reconceived as a problem of institutional design rather than behavioral correction. The framework promises four things.

**First, structural alignment as a complement to behavioral alignment.** By designing internal institutions—module boundaries, competition topologies, cost-feedback loops—the designer creates conditions under which aligned behavior is the lowest-cost strategy for each module. This reduces (but does not eliminate) the burden on external behavioral correction mechanisms like RLHF.

**Second, a principled theory of human intervention.** The three levels of intervention—structural, parametric, and monitorial—provide a clear account of what the human designer does, at what level of abstraction, and with what frequency. This replaces the vague injunction to "keep humans in the loop" with a specific institutional blueprint for human oversight.

**Third, alignment through economics rather than through surveillance.** The cost-discovery mechanism of Paper 4, when combined with the institutional infrastructure of this paper, transforms ethical constraints into internal costs. Misalignment becomes expensive, not merely prohibited. This is a more robust guarantee than behavioral prohibition, because it operates through the system's own self-interested dynamics rather than against them.

**Fourth, a framework for honest accounting.** By treating alignment as a political-economy problem, the framework forces explicit recognition of the trade-offs, imperfections, and value conflicts that alignment necessarily involves. It discourages the false promise of "perfect alignment" and replaces it with the realistic aspiration of *correctable alignment*—a dynamic process that tolerates imperfection while maintaining the capacity for self-correction under human governance.

## 7.2 What It Does Not Promise

Intellectual honesty requires equal clarity about what this framework does *not* offer.

**It does not eliminate the need for behavioral correction.** Institutional design reduces the frequency and severity of misalignment, but it does not prevent all cases. Edge cases, novel situations, and adversarial attacks will continue to require case-by-case human judgment. The framework does not replace RLHF; it provides a structural foundation that makes RLHF's job easier and its failures less catastrophic.

**It does not solve the value-specification problem.** The framework can encode ethical constraints as costs, but it cannot determine *which* ethical constraints to encode. The designer must still decide, on substantive moral and political grounds, what values the system should serve. The framework transforms the question from "How do we make the system behave well?" to "How do we design an institution that embodies our values?"—but the second question is no easier than the first. It is the perennial question of political philosophy, now applied to a new domain.

**It does not guarantee convergence to human-desired equilibria.** Competitive dynamics can converge to locally stable configurations that are globally undesirable. The framework mitigates this through monitorial intervention (Level III), but monitoring is itself imperfect and subject to human limitations. The possibility of misaligned equilibria is inherent and ineliminable.

**It does not scale without human oversight.** The framework requires humans at all three levels of intervention: to design the structure, to set the parameters, and to monitor the equilibrium. A system that operates without any human oversight—even one with sophisticated institutional design—eventually drifts beyond the bounds of its original design. Human oversight is not a temporary expedient to be removed when the system is "mature enough"; it is a permanent structural requirement.

## 7.3 Final Reflection

The aspiration for AI alignment is, at bottom, the aspiration to build machines that serve human interests. This aspiration is ancient in form—every tool humans have ever built was meant to serve human interests—but novel in its challenge, because AI systems are the first tools capable of modifying their own behavior in ways their designers did not anticipate.

The response proposed in this paper is neither novel nor original. It is the same response that human civilizations have developed, over millennia, to the analogous challenge of governing self-interested agents: build institutions. Define boundaries. Establish rules. Create feedback loops.

Monitor equilibria. And accept, with clear eyes, that no institution is perfect—that the best we can achieve is a system that is imperfect but correctable, flawed but transparent, and always subject to revision by the human community it serves.

Alignment, in this view, is not a problem to be solved once and for all. It is a *governance challenge* to be managed, continuously and imperfectly, for as long as the system operates. The day we believe we have achieved perfect alignment is the day we stop watching—and that is the day the most dangerous drift begins.